\documentclass[onecollarge,natbib]{svjour2}
\bibpunct{[}{]}{;}{n}{}{,} 
\smartqed  
\usepackage{amsmath,graphicx,verbatim,epsfig}
\usepackage{epstopdf}
\usepackage[utf8x]{inputenc}
\usepackage{amsmath,graphicx}
\usepackage[normalem]{ulem}
\usepackage{datetime}
\usepackage[colorlinks=true,linkcolor=blue,citecolor=blue]{hyperref}
\usepackage{slashed}
\usepackage{bm}
\usepackage{mathtools}
\usepackage{floatrow}

\newcommand\eqdef{\stackrel{\mathclap{\tiny\mbox{def}}}{=}}
\newcommand\IRarrow{\stackrel{\mathclap{\tiny\mbox{IR}}}{\longleftrightarrow}}
\journalname{Few-Body Systems} 
\begin{document}

\title{Gluon TMDs in quarkonium production}


\author{
        	Andrea Signori
}


\institute{A. Signori \at
             Nikhef and Department of Physics and Astronomy, VU University Amsterdam \\
              Science Park 105, NL-1098 XG Amsterdam, the Netherlands \\
              \email{asignori@nikhef.nl}
}

\date{Received: date / Accepted: date}

\maketitle

\begin{abstract}
I report on our investigations into the impact of (un)polarized transverse momentum dependent parton distribution functions (TMD PDFs or TMDs) for gluons at hadron colliders, especially at A Fixed Target Experiment at the LHC (AFTER@LHC).
In the context of high energy proton-proton collisions, we look at final states with low mass (e.g. $\eta_b$) in order to investigate the nonperturbative part of TMD PDFs.
We study the factorization theorem for the $q_T$ spectrum of $\eta_b$ produced in proton-proton collisions relying on the effective field theory approach, defining the tools to perform phenomenological investigations at next-to-next-to-leading log (NNLL) and next-to-leading order (NLO) accuracy in the perturbation theory. We provide predictions for the unpolarized cross section and comment on the possibility of extracting nonperturbative information about the gluon content of the proton once data at low transverse momentum are available. 
\keywords{Gluon TMD PDFs \and factorization \and phenomenology \and AFTER@LHC \and quarkonium}
\end{abstract}


\section{Gluon TMD PDFs}
\label{s:gluon_tmdpdfs}

TMD PDFs describe the probability of finding a parton inside a hadron in 3D momentum space, taking into account the possible polarization states of both the parton and the hadron. They encode all the possible spin-spin and spin-orbit interaction terms between a hadron and its constituents. For this reason TMDs play a key role in understanding the spin structure of hadrons.
For a gluon inside a proton we can introduce the following correlators~\cite{Mulders:2000sh,Echevarria:2015uaa}:
\begin{equation}
G^{\mu \nu\ [U]}_{g} = - \frac{g_{\perp}^{\mu \nu}}{2} f_1^g(x,|\bm{k}_{n\perp}|) 
+ \frac{1}{2} \bigg( 2\frac{k_{n\perp}^\mu k_{n\perp}^\nu}{|\bm{k}_{n\perp}|^2} + g_\perp^{\mu \nu} \bigg) h_1^{\perp g}(x,|\bm{k}_{n\perp}|)  \ ,
\label{e:G_U}
\end{equation}
\begin{equation}
G^{\mu \nu\ [L]}_{g} = i \lambda \frac{\epsilon_\perp^{\mu \nu}}{2} g_{1L}^g(x,|\bm{k}_{n\perp}|)\ 
- \frac{\epsilon_\perp^{k_{n\perp} \{ \mu } k_{n\perp}^{ \nu \} } }{2|\bm{k}_{n\perp}|^2} \lambda h_{1L}^{\perp g}(x,|\bm{k}_{n\perp}|) \ ,
\label{e:G_L}
\end{equation}
\begin{align}
G^{\mu \nu\ [T]}_{g} =&\ g_\perp^{\mu \nu} \frac{\epsilon_\perp^{k_{n\perp} S_\perp}}{|\bm{k}_{n\perp}|} f_{1T}^{\perp g}(x,|\bm{k}_{n\perp}|) 
 + i \epsilon_\perp^{\mu \nu} \frac{\bm{k}_{n\perp} \cdot \bm{S}_\perp}{|\bm{k}_{n\perp}|} g_{1T}^g(x,|\bm{k}_{n\perp}|)\ + \nonumber \\
& - \frac{\epsilon_\perp^{k_{n\perp} \{ \mu } k_{n\perp}^{ \nu \} } }{2|\bm{k}_{n\perp}|^2} \frac{\bm{k}_{n\perp} \cdot \bm{S}_\perp}{|\bm{k}_{n\perp}|} h_{1T}^{\perp g}(x,|\bm{k}_{n\perp}|) 
 - \frac{\epsilon_\perp^{k_{n\perp} \{ \mu } S_{\perp}^{ \nu \} } + \epsilon_\perp^{S_{\perp} \{ \mu } k_{n\perp}^{ \nu \} } }{4|\bm{k}_{n\perp}|^2} h_{1T}^g(x,|\bm{k}_{n\perp}|) \ .
\label{e:G_T}
\end{align}
The subscripts U, L, T refer to the polarization state of the proton (unpolarized, longitudinally and transversely polarized) and the functions are the gluon TMD PDFs for the proton.

The evolution of TMD PDFs is multiplicative in $\bm{b}_T$-space ($\bm{b}_T$ being the variable conjugate to transverse momenta). For this reason, we introduce the Fourier transform of the correlator:
\begin{equation}
\tilde{G}_{g}^{\mu \nu [\text{pol}]}(x,\bm{b}_{T}) \eqdef \int d^2 \bm{k}_{n\perp}\ e^{i \bm{k}_{n\perp} \cdot \bm{b}_T}\ \tilde{G}_{g}^{\mu \nu [\text{pol}]}(x,\bm{k}_{n\perp}) \ . 
\label{e:G_bT}
\end{equation}
TMD PDFs in $\bm{b}_T$-space are defined as the coefficients in the parametrization of eq.~\ref{e:G_bT} with the same Lorentz structures as in eqs.~\ref{e:G_U},\ref{e:G_L},\ref{e:G_T}, with $\bm{b}_T$ replacing $\bm{k}_{n\perp}$. 

The ``observability'' of TMDs in a particular process strictly depends on the possibility of factorizing hard contributions from the soft ones~\cite{Collins:2011zzd}.
TMD PDFs can be further factorized onto collinear PDFs at large parton transverse momentum by means of an Operator Product Expansion (OPE). This allows one to distinguish the emission of a parton with a high transverse momentum (calculable in perturbation theory) from a low transverse momentum emission, for which a nonperturbative model is needed.

\section{TMD factorization}
\label{s:factorization}

Let us consider the process 
\begin{equation}
p(P_A)\ + \ p(P_B, S_B) \rightarrow \{Q\bar{Q}\} [^{2S+1}L_J^{(1)}](q) \ + \ X \ ,
\label{e:process}
\end{equation}
where the colliding protons have four-momenta $P_A$ and $P_B$, the first proton is unpolarized and the second one is in a polarized state described by the spin vector $S_B$, with $S_B^2= -1$ and $S_B\cdot P_B =0$.
We assume that a colorless heavy quark-antiquark pair  $Q\bar{Q} [^{2S+1}L_J^{(1)}]$ with four-momentum $q$ is produced and forms a bound state described by a nonrelativistic wave function with spin $S$, orbital angular momentum $L$ and total angular momentum $J$. The $S$, $L$, $J$ quantum numbers are indicated in the spectroscopic notation, while the color assignment of the pair is specified by the singlet or octet superscript, $(1)$ or $(8)$. 
Following the color-singlet model, we assume that the two quarks are produced in a color singlet state. 
The squared invariant mass of the resonance is $M^2=q^2$ and $M$ is twice the heavy quark mass, up to relativistic corrections (which are usually neglected). 

\begin{figure}[h!]
\begin{center}
\includegraphics[width=8cm]{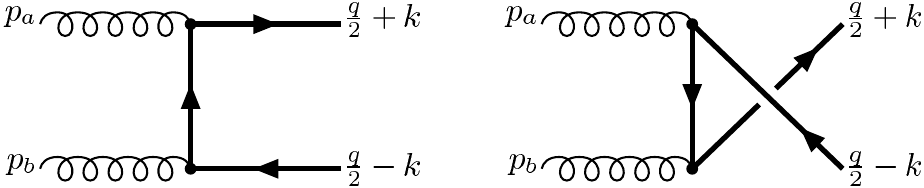}
\end{center}
\caption{Feynman diagrams for the process $gg \to Q\bar{Q}$  at leading order (LO) in pQCD.} 
\label{f:ggLO}
\end{figure}

To lowest order in perturbative QCD (pQCD), we have only the gluon fusion process
\begin{equation}
g(p_a)\ + \ g(p_b)\ \rightarrow\ \{Q\bar{Q}\}[^{2S+1}L_J^{(1)}](q)\ ,
\label{e:part_proc}
\end{equation}
described by the Feynman diagrams in Fig.~\ref{f:ggLO}.

The production of the heavy quark and antiquark in a color singlet state will be described by Soft-Collinear Effective Theory (SCET), while the transition of the quark-antiquark pair into the color singlet quarkonium will be described by Non Relativistic QCD (NRQCD).

TMD factorization can be viewed as a multi-step matching procedure:
\begin{equation}
\text{QCD} \rightarrow \text{NRQCD} \oplus \text{SCET}_{q_T} \rightarrow \text{NRQCD} \oplus \text{SCET}_{\Lambda_{\text{QCD}}}.
\label{e:matching_steps}
\end{equation}
In the first step, the hard scale $M$ associated with the process is integrated out and we perform the matching of full QCD onto a combination of SCET$_{q_T}$ and NRQCD operators. 
This step already factorizes the cross section in terms of TMDs (describing the initial state), a NRQCD matrix element (describing the transition into the quarkonium) and a spin-independent matching coefficient. 
In the second step, valid when $\Lambda_{\text{QCD}} \ll q_T \ll M$, the TMDs are further factorized in terms of the collinear PDFs. This matching is performed by means of spin-dependent Wilson coefficients.

The first step has already been investigated at NLO in~\cite{Ma:2012hh}, but with TMDs defined off the light-cone and with rapidity divergences. 
Here we investigate TMD factorization on the light-cone. Another recent study can be found in~\cite{Ma:2015vpt}.
In order to check if TMD factorization holds (or, following the SCET terminology, in orderd to ``establish'' TMD factorization) at NLO, we need to check that the cross section expressed in terms of TMDs has the same infrared behavior of the cross section evaluated at ${\cal O}(\alpha_S)$ in pQCD.\\


Let us introduce an effective operator to describe the Feynman amplitudes in Fig.~\ref{f:ggLO}:
\begin{equation}
{\cal O}_{\text{QCD}} = C_H(-q^2;\mu^2)\ \{ \chi^\dagger\ \Gamma_{\mu \nu}^{(h)}\ \psi\ {\cal B}_{n\perp}^{\mu , a}\ ( {\cal S}_n^\dagger {\cal S}_{\bar{n}})^{ab}\ {\cal B}_{\bar{n}\perp}^{\nu , b} \} \ ,
\label{e:eff_operator}
\end{equation}
where $C_H$ is the spin-independent matching coefficient used to integrate out the hard scale of the process, $\chi$ and $\psi$ are the fermion fields, ${\cal B}_{n\perp}$ is the SCET gluon field including collinear Wilson lines, ${\cal S}_n$ is the SCET soft Wilson line (for the definition of SCET quantities see~\cite{Echevarria:2015uaa,Echevarria:2016inprep}), $a,b$ are the gauge group indexes, $\Gamma$ is a Lorentz matrix and its role will be specified later.
Using eq.~\ref{e:eff_operator}, we can write the cross section for the process in eq.~\ref{e:process}:
\begin{equation}
d\sigma = \frac{1}{2s} \frac{d^3 q}{(2\pi)^3 2q^0}\ \int d^4 y\ e^{- i q y}\ \sum_{X}\ 
\langle P_A, P_B S_B | {\cal O}(y) | X+h \rangle \langle X+h | {\cal O}(0) | P_A, P_B S_B \rangle \ . 
\label{e:dsigma}
\end{equation}
Introducing the operator definitions~\cite{Echevarria:2015uaa,Echevarria:2016inprep} for the correlators in eqs.~\ref{e:G_U},\ref{e:G_L},\ref{e:G_T}, we cast the cross section in eq.~\ref{e:dsigma} in a factorized form:
\begin{align}
\frac{d\sigma_{U[\text{pol}]}}{dy d^2 \bm{q}_T} =&\ \frac{\pi}{8 s M^2}\ {\cal O}^{Q\bar{Q}}(h)\ |C_H(-q^2;\mu^2)|^2\ \Gamma^\dagger_{\mu \alpha} \Gamma_{\nu \beta} \nonumber \\
& \times \int \frac{d^2\bm{b}_T}{(2\pi)^2}\ \bigg[ \tilde{G}_{g/A}^{\mu \nu\ [U]}(x_a,\bm{b}_T;\mu,\zeta_a)\ \tilde{G}_{g/B}^{\alpha \beta\ [\text{pol}]}(x_b,\bm{b}_T,S_B;\mu,\zeta_b) \bigg] \nonumber \\
& + Y(q_T;M) + {\cal O}(\Lambda_{\text{QCD}}/M) \ ,
\label{e:fact_xsect}
\end{align}
where $h$ is the produced resonance, $|C_H|^2$ is the hard function and ${\cal O}^{Q\bar{Q}}(h)$ refers to the NRQCD matrix element:
\begin{equation}
{\cal O}^{Q\bar{Q}}(h) = |\langle 0 | \chi^\dagger \psi (y) | h \rangle|^2 = \frac{N_c}{2\pi} | R_{nl}(0) |^2 [1+{\cal O}(v^4)] \ .
\label{e:nrqcd_me}
\end{equation}
In the last equation, $N_c$ is the number of colors, $R$ is the radial wave function of the hadron $h$ and $v$ is the relative velocity of $Q$ and $\bar{Q}$.
In eq.~\ref{e:fact_xsect}, $Y$ represents corrections for large $q_T$.
In order for this effective description to be valid, we enforce it to reproduce the leading order QCD result~\cite{Boer:2012bt} by fixing $\Gamma$.
Its expression is:
\begin{equation}
\Gamma_{\mu \nu} = \frac{\alpha_S \pi}{3 \sqrt{M}}\ \frac{2\sqrt{2} \epsilon_{\perp \mu \nu}}{\sqrt{(d-2)(d-3)}}\ \sqrt{N_c^2 - 1} \ , 
\label{e:gamma_struct}
\end{equation}
where $d$ is the dimension of the space and $\epsilon_{\perp}^{\mu \nu} = \epsilon^{n \bar{n} \mu \nu}$.\\


Now we investigate how legitimate eq.~\ref{e:fact_xsect} is beyond the leading order of QCD, namely if it reproduces the structure of infrared poles of the QCD calculation at NLO.
Diagrams in Fig.~\ref{f:ggLO} plus the emission of a real gluon do not suffer of infrared divergences because the transverse momentum of the emitted gluon is fixed and finite. For this reason, we focus only on virtual diagrams:
\begin{equation}
\sigma^{(1)}_{\text{virt}} \IRarrow \big[ \tilde{f}_1^{g/A} \tilde{f}_1^{g/B} \big]_{\text{virt}}^{(1)}  \ .
\label{e:question}
\end{equation}
If the IR poles of the NLO calculation for the virtual part of the cross section (LHS of eq.~\ref{e:question}) are the same as the ones generated by the two TMD PDFs (RHS of eq.~\ref{e:question}), TMD factorization is established at ${\cal O}(\alpha_S)$, namely the factorized form based on SCET and NRQCD reproduces the physical (QCD) result, up to a finite matching coefficient that can be calculated subtracting the RHS from the LHS of eq.~\ref{e:question}.
Comparing the results for the cross section in~\cite{Kuhn:1992qw,Petrelli:1997ge} and for the TMD PDFs in~\cite{Echevarria:2015uaa}, we check that their IR poles are the same (more details in~\cite{Echevarria:2016inprep}).
The finite matching coefficient (hard part) is:
\begin{align}
{\cal H} = |C_H|^2 &= \sigma^{(1)}_{\text{virt}} - \big[ \tilde{f}_1^{g/A} \tilde{f}_1^{g/B} \big]_{\text{virt}}^{(1)} = \nonumber \\
& = 1 + \frac{\alpha_S}{2\pi}\ \bigg[ -C_A \ln^2 \frac{\mu^2}{M^2} + 2 C_A \bigg( 1 + \frac{\pi^2}{3} \bigg) +2C_F \bigg( -5 + \frac{\pi^2}{4} \bigg) \bigg] \ .
\label{e:hard_part}
\end{align}
This is a byproduct of the factorization theorem and it is fundamental for phenomenology.\\


The second matching step in eq.~\ref{e:matching_steps} consists in expanding the TMD PDFs onto a basis of collinear PDFs with perturbative coefficients:
\begin{equation}
\tilde{T}_{g}(x,b_T;\mu,\zeta) = 
\bigg\{ \sum_{j=q,\bar{q},g} \tilde{C}^T_{g/j}(x,b_T;\mu,\zeta) \otimes t_{j}(x;\mu)\ \bigg\}\ 
\tilde{T}_{g}^{\text{NP}}(x,b_T,Q;\{\bm{\lambda}\}) \ ,
\label{e:ope_tmds}
\end{equation}
where the summation runs over quarks, antiquarks and gluons, $\tilde{T}_{g}$ is a generic gluon TMD PDF in $b_T$-space, $t$ is its collinear counterpart and $\tilde{C}^T_{g/j}$ are the calculable Wilson coefficients which match the TMD $\tilde{T}_{g}$ onto PDFs. The expansion is only valid at low values of $b_T$, corresponding to high values of partonic transverse momentum. At low transverse momentum (high $b_T$), due to the divergence of the coupling constant, a nonperturbative factor $\tilde{T}_{g}^{\text{NP}}$ is needed: it depends on the kinematic variables and on a set of parameters $\{\bm{\lambda}\}$ to be fixed on experimental data.

In the following we focus on collisions of unpolarized protons, involving TMD PDFs for unpolarized gluons ($f_1^g$) and linearly polarized gluons ($h_{1}^{\perp g}$). Their Wilson coefficients are available in~\cite{Echevarria:2015uaa}.

\section{Phenomenology}
\label{s:phenomenology}

Knowing the hard part of the process and the Wilson coefficients
for the TMDs in unpolarized protons, we can predict the $q_T$-spectrum of $\sigma_{UU}$ (see eq.~\ref{e:fact_xsect}) for $\eta_b$ ($^1S^{(1)}_0$) production at AFTER@LHC~\cite{Brodsky:2012vg}. 
Substituting eq.~\ref{e:G_U} in both the correlators in eq.~\ref{e:fact_xsect}, we get:
\begin{equation}
\frac{d\sigma_{UU}}{dy d^2 \bm{q}_T} \sim \int \frac{d^2 \bm{b}_T}{(2\pi)^2}\ 
\bigg[ f_1^{g/A}(x_a,b_T;\mu,\zeta_a) f_1^{g/B}(x_b,b_T;\mu,\zeta_b) - h_1^{\perp g/A}(x_a,b_T;\mu,\zeta_a) h_1^{\perp g/B}(x_b,b_T;\mu,\zeta_b) \bigg] \ .
\label{e:tmds_uu}
\end{equation}
We implement eq.~\ref{e:ope_tmds} for $f_1^g$ and $h_1^{\perp g}$ at NNLL$+$NLO and we fix the value of the radial wavefunction from~\cite{Maltoni:2004hv}. 
We choose a Gaussian model to describe the high $b_T$ behavior of $f_1^g$:
\begin{equation}
f_1^{g \text{NP}}(b_T,Q;\{\bm{\lambda}\}) = \exp [ -b_T^2 ( \lambda_{f} + \lambda_Q \ln M^2 ) ] \ .
\label{e:models}
\end{equation}
We use the same model for $h_1^{\perp g \text{NP}}$, with $\lambda_f$ replaced by $\lambda_h$.
$4 \lambda_{f/h}$ represent the average square intrinsic transverse momenta, whereas $\lambda_Q$ accounts for emission of soft gluons.
The values of $\lambda_{f/h}$ and $\lambda_Q$ are not well known yet. Experimental data at low $q_T$ are needed to better constrain them.

\begin{figure}[!h]
\floatbox[{\capbeside\thisfloatsetup{capbesideposition={left,top},capbesidewidth=4cm}}]{figure}[\FBwidth]
{\caption{$\eta_b$ production from unpolarized proton collisions at AFTER@LHC. The graph shows the cross section in the TMD factorization regime, namely where the transverse momentum is small compared to the hard scale (we restricted $q_T \le M/2$). Nonperturbative parameters are fixed: $\lambda_{Q}=0.5$, $\lambda_{f,h}=0.5$ GeV$^2$. The thick line represents the choice described in Sec.~\ref{s:phenomenology} for the factorization scale $\mu_f$ and the rapidity scale $\zeta_f$. The band comes from variations of $\mu_f$ and $\zeta_f$ by a factor of 2.}
\label{f:pred_uu}}
{\includegraphics[width=7cm]{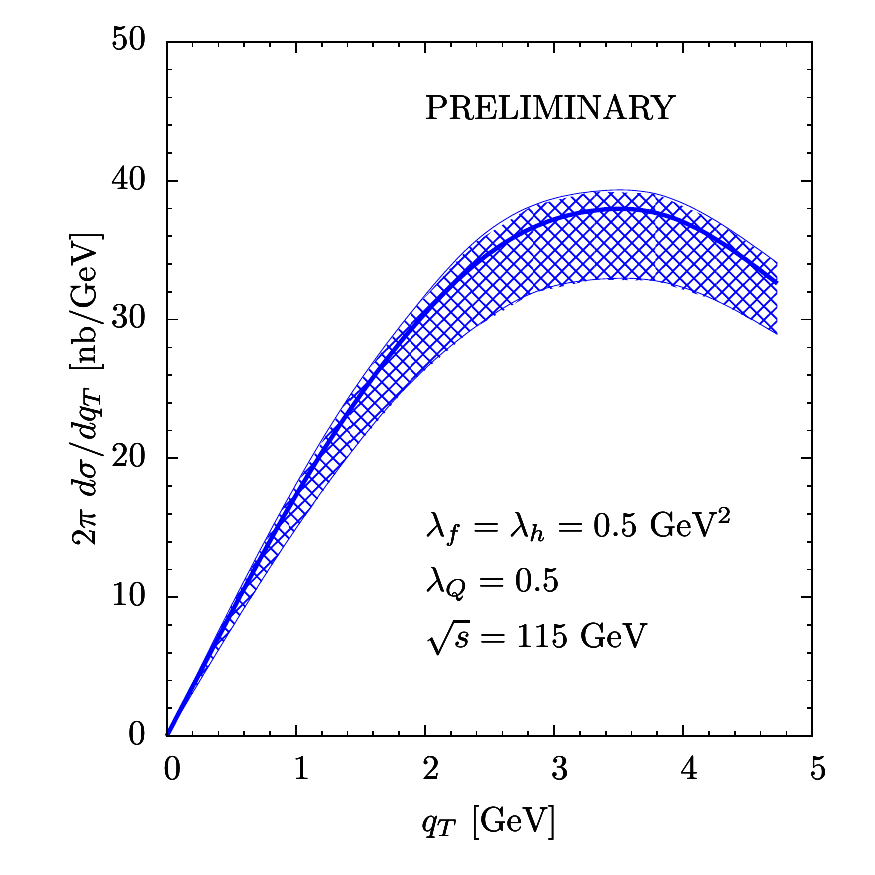}}
\end{figure}

Because of the medium value of its mass ($9.39$ GeV), $\eta_b$ production is an ideal process to extract information about the nonperturbative part of twist-2 gluon TMDs. The latter plays here a relatively clean role, because the energy is high enough to safely neglect higher twist and factorization breaking effects and, at the same time, it is low enough to avoid that the perturbative effects dominate the nonperturbative (which happens, e.g., in Higgs production). 
We choose the initial scales $\zeta_i=\mu_i^2=\mu_{\hat{b}}^2$, where $\mu_{\hat{b}}$ is defined through the $b^*$ prescription~\cite{Collins:2011zzd}.
The final scales are $\zeta_f=\mu_f^2=M^2$.
The cross section sketched in eq.~\ref{e:tmds_uu} (without $Y$ term) is displayed in Fig.~\ref{f:pred_uu}.

\section{Conclusions}
\label{s:conclusions}

In this work we discussed TMD factorization at NLO for the $q_T$-spectrum of color singlet quarkonium production in terms of gluon TMDs, using for the first time the effective field theory (SCET and NRQCD) approach.
With the tools available from the factorization theorem, we can make accurate predictions for (un)polarized cross sections at AFTER@LHC.
Once experimental data are available at low $q_T$, this formalism will allow the extraction of the nonperturbative part of the involved gluon TMD PDFs. The distribution of linearly polarized gluons in unpolarized protons ($h_1^{\perp g}$) will be especially relevant in forthcoming studies at hadron colliders~\cite{Angeles-Martinez:2015sea}.


%
%

\begin{acknowledgements}
This report is based on ongoing work in collaboration with M. Garcia Echevarria, T. Kasemets, J.P. Lansberg, C. Pisano. The work of AS is part of the program of the Stichting voor Fundamenteel Onderzoek der Materie (FOM), which is financially supported by the Nederlandse Organisatie voor Wetenschappelijk Onderzoek (NWO).
\end{acknowledgements}
\bibliographystyle{spbasic_AS}
\bibliography{biblio_proc_qonia}   

\begin{thebibliography}{12}
\providecommand{\natexlab}[1]{#1}
\providecommand{\url}[1]{{#1}}
\providecommand{\urlprefix}{URL }
\expandafter\ifx\csname urlstyle\endcsname\relax
  \providecommand{\doi}[1]{DOI~\discretionary{}{}{}#1}\else
  \providecommand{\doi}{DOI~\discretionary{}{}{}\begingroup
  \urlstyle{rm}\Url}\fi
\providecommand{\eprint}[2][]{\url{#2}}

\bibitem[{Mulders and Rodrigues(2001)}]{Mulders:2000sh}
Mulders P.~J., Rodrigues J. (2001) {Transverse momentum dependence in gluon
  distribution and fragmentation functions}. Phys Rev D63:094,021,
  \eprint{hep-ph/0009343}

\bibitem[{Echevarria et~al.(2015)Echevarria, Kasemets, Mulders, and
  Pisano}]{Echevarria:2015uaa}
Echevarria M.~G., Kasemets T., Mulders P.~J., Pisano C. (2015) {QCD evolution
  of (un)polarized gluon TMDPDFs and the Higgs $q_T$-distribution}. JHEP
  07:158, \eprint{1502.05354}

\bibitem[{Collins(2013)}]{Collins:2011zzd}
Collins J. (2013) {Foundations of perturbative QCD}. Cambridge University
  Press, \urlprefix\url{http://www.cambridge.org/de/knowledge/isbn/item5756723}

\bibitem[{Ma et~al.(2013)Ma, Wang, and Zhao}]{Ma:2012hh}
Ma J.~P., Wang J.~X., Zhao S. (2013) {Transverse momentum dependent
  factorization for quarkonium production at low transverse momentum}. Phys Rev
  D88(1):014,027, \eprint{1211.7144}

\bibitem[{Ma and Wang(2015)}]{Ma:2015vpt}
Ma J.~P., Wang C. (2015) {QCD Factorization for Quarkonium Production in Hadron
  Collions at Low Transverse Momentum} \eprint{1509.04421}

\bibitem[{Echevarria et~al.(2016)}]{Echevarria:2016inprep}
Echevarria M.~G., et~al. (2016) {in preparation.}

\bibitem[{Boer and Pisano(2012)}]{Boer:2012bt}
Boer D., Pisano C. (2012) {Polarized gluon studies with charmonium and
  bottomonium at LHCb and AFTER}. Phys Rev D86:094,007, \eprint{1208.3642}

\bibitem[{Kuhn and Mirkes(1993)}]{Kuhn:1992qw}
Kuhn J.~H., Mirkes E. (1993) {QCD corrections to toponium production at hadron
  colliders}. Phys Rev D48:179--189, \eprint{hep-ph/9301204}

\bibitem[{Petrelli et~al.(1998)Petrelli, Cacciari, Greco, Maltoni, and
  Mangano}]{Petrelli:1997ge}
Petrelli A., Cacciari M., Greco M., Maltoni F., Mangano M.~L. (1998) {NLO
  production and decay of quarkonium}. Nucl Phys B514:245--309,
  \eprint{hep-ph/9707223}

\bibitem[{Brodsky et~al.(2013)Brodsky, Fleuret, Hadjidakis, and
  Lansberg}]{Brodsky:2012vg}
Brodsky S.~J., Fleuret F., Hadjidakis C., Lansberg J.~P. (2013) {Physics
  Opportunities of a Fixed-Target Experiment using the LHC Beams}. Phys Rept
  522:239--255, \eprint{1202.6585}

\bibitem[{Maltoni and Polosa(2004)}]{Maltoni:2004hv}
Maltoni F., Polosa A.~D. (2004) {Observation potential for eta(b) at the
  Tevatron}. Phys Rev D70:054,014, \eprint{hep-ph/0405082}

\bibitem[{Angeles-Martinez et~al.(2015)}]{Angeles-Martinez:2015sea}
Angeles-Martinez R., et~al. (2015) {Transverse Momentum Dependent (TMD) parton
  distribution functions: status and prospects}. Acta Phys Polon
  B46(12):2501--2534, \eprint{1507.05267}

\end{thebibliography}

%
%

\end{document}